\documentclass[letterpaper,journal]{IEEEtran}
\usepackage{amsmath,amsfonts,amssymb}
\usepackage{array}
\usepackage{booktabs}
\usepackage{multirow}
\usepackage{graphicx}
\usepackage{textcomp}
\usepackage{enumitem}
\usepackage{stfloats}
\usepackage{url}
\usepackage{cite}
\usepackage{xcolor}
\usepackage[hidelinks]{hyperref}

\newcommand{\psok}{\texttt{<psok>}}
\newcommand{\peok}{\texttt{<peok>}}
\newcommand{\rej}{\texttt{<reject/>}}

\begin{document}
\raggedbottom

\title{WakeCue-ASR: A Speech-LLM Framework for Enrollment-Free Wake-Word-Guided Target-Speaker Recognition in 
Complex Noisy Scenarios for Seamless Human-Machine Interaction}

\author{Yuxiang~Mei, Yuchen~Yan, Dongxing~Xu, Yu~Gao, and~Yanhua~Long, Member, IEEE
\thanks{Y.~Mei and Y.~Long are with Shanghai Normal University, Shanghai, China
(corresponding author: Yanhua Long). (e-mail: m153517@icloud.com; yanhua@shnu.edu.cn). }
\thanks{Yuchen Yan is with Baidu, Inc. (e-mail: yanyuchen01@baidu.com).}
\thanks{Dongxing Xu is with Unisound AI Technology Co., Ltd., Beijing, China (e-mail: xudongxing@unisound.com).}
\thanks{Yu Gao is with Midea AI Research Center (e-mail: gaoyu11@midea.com).}
\thanks{This work was done while Yuxiang Mei was an intern at Baidu.}}

\markboth{}%
{Mei \MakeLowercase{\textit{et al.}}: Speech-LLM for Wake-Word-Guided Target-Speaker Recognition}

\maketitle

\begin{abstract}
Target-speaker automatic speech recognition is essential for reliable human-machine 
interaction in complex acoustic environments, where overlapping speech and background noise 
make it difficult to identify and recognize the intended speaker. Existing approaches typically 
rely on speaker enrollment or cascaded processing pipelines, limiting their flexibility in practical applications.
To address these limitations, we propose WakeCue-ASR, an enrollment-free Speech-LLM framework 
that requires neither enrollment speech nor an explicit voiceprint module, instead leveraging 
the wake-word text as a text-cue to guide target-speaker recognition.
Specifically, a text-cue encoder extracts lexical representations from the wake word, 
which are integrated through cross-attention into a speech encoder to automatically 
and implicitly track the voiceprint of the speaker who triggered the wake word, thereby 
establishing an implicit association between the wake-word occurrence and the target speaker.
This enables the model to selectively recognize only the subsequent commands from the 
wake-word speaker without requiring enrollment speech or explicit wake-word boundaries.
Furthermore, a large language model (LLM) unifies command transcription and non-target speaker 
rejection within a single autoregressive framework. 
To improve robustness against complex noisy acoustic interference, we propose a large-scale 
training data synthesis and simulation pipeline with a progressive four-stage training strategy.
Experimental results on multiple real-world industrial test sets demonstrate the effectiveness 
of the proposed framework; notably, on a complex smart-home test set (Midea AI Challenge),
our model achieves a command character error rate (CER) of 16.15\% and a non-target speaker
rejection rate of 91.56\%. 
These results highlight the potential of wake-word-guided implicit voiceprint tracking, 
and of integrating target-speaker recognition and rejection within a unified Speech-LLM framework 
for robust and seamless human-machine interaction.
Our code\footnote{\url{https://github.com/YuCeong-May/WakeCue-ASR}} and 
demo\footnote{\url{https://huggingface.co/spaces/YuCeong-May/WakeCue-ASR}} are publicly available.

\end{abstract}


\section{Introduction}
\label{sec:introduction}

\IEEEPARstart{V}{oice}-based human-machine interaction has become increasingly prevalent 
in smart homes and intelligent devices, enabling users to control appliances through natural 
spoken commands. A typical interaction begins with a predefined wake word, followed by a 
command that specifies the intended action. However, real-world acoustic environments 
often contain overlapping speech from multiple speakers, background conversations, and 
noise generated by household appliances. Under such conditions, a voice-controlled system
must not only recognize the spoken command but also determine whether it originates from 
the speaker who initiated the interaction. Commands issued by other speakers should be 
ignored, even when they are clearly recognizable. This requirement makes target-speaker 
automatic speech recognition (TS-ASR) a fundamental component of reliable and seamless 
human-machine interaction.

Conventional TS-ASR approaches typically identify the target speaker using an enrollment utterance, 
from which speaker representations are extracted to guide speech separation or 
recognition~\cite{delcroix2018single,zmolikova2019speakerbeam,kanda2019auxiliary}. Although 
effective in controlled scenarios, enrollment-based methods require a prior speaker enrollment step, 
introducing additional interaction requirements and dependence on the quality of the enrollment speech.
These limitations are particularly relevant to shared household devices, where multiple users 
may interact with the same system without prior speaker registration. To alleviate this dependency, 
anchored speech recognition~\cite{wang2019anchored,raj2023anchored} uses a short speech segment 
from the current interaction as an implicit speaker reference. Nevertheless, wake-word segments 
are typically brief and may themselves be contaminated by competing speech and environmental noise, 
making reliable speaker representation extraction challenging. Furthermore, cascaded TS-ASR systems that 
combine speaker verification, speech separation, and automatic speech recognition (ASR) 
introduce large additional computational complexity and may suffer from error propagation across 
independently optimized modules.

An alternative perspective is to exploit the lexical information already available 
during wake-word-based interaction, thereby enabling target-speaker ASR to be achieved 
implicitly in a seamless interaction manner. Unlike conventional enrollment speech, 
the wake-word text is predefined and naturally associated with the speaker who initiates the interaction. 
This provides a distinctive form of target-speaker cue: rather than specifying the target through 
a separately recorded enrollment speech, the system can identify the speaker by locating the 
occurrence of a known lexical sequence within the observed speech mixture. 
Once the wake-word occurrence is identified, its associated acoustic characteristics 
can potentially guide the recognition of subsequent commands from the same speaker. 
Importantly, this formulation does not require an explicitly segmented wake-word recording 
or a previously enrolled speaker identity. It therefore offers a natural foundation for 
enrollment-free target-speaker recognition in wake-word-driven applications.

Recent advances in speech large language models (Speech-LLMs) have demonstrated the potential 
of integrating pretrained speech encoders with large language models for speech recognition and audio-language 
understanding~\cite{xu2025qwen3omni,shi2026qwen3asr,tang2024salmonn,ma2024embarrassingly}. By projecting acoustic 
representations into the embedding space of a language model, these architectures benefit from pretrained linguistic knowledge 
while supporting flexible instruction-based decoding. However, conventional Speech-LLM architectures are not explicitly 
designed to associate a lexical cue with a particular speaker in overlapping speech. Moreover, target-speaker recognition 
requires an additional decision beyond transcription: \textit{when the wake-word speaker is absent from the command 
segment, the system must reject the input rather than generate a potentially incorrect command}. 
Integrating speaker selection, command transcription, and rejection into a unified Speech-LLM framework 
therefore presents an important research challenge.

To address these challenges, we propose WakeCue-ASR, 
an enrollment-free wake-word-guided Speech-LLM framework 
for target-speaker recognition in complex acoustic environments. The proposed approach 
introduces a text-cue encoder that transforms the known wake-word text into contextual 
representations, which are subsequently integrated into a speech encoder 
through cross-attention. This design enables the acoustic encoder to condition its 
representations on the lexical cue and learn speaker-selective features associated 
with the wake-word occurrence. The resulting acoustic representations are projected into 
the embedding space of a pretrained large language model, which is adapted using low-rank adaptation 
(LoRA)~\cite{qwen2025qwen25,hu2022lora}. Through a unified autoregressive decoding process, 
the model generates either the command spoken by the target speaker or an explicit rejection 
token when the target speaker is absent. Unlike conventional cascaded approaches, 
the proposed WakeCue-ASR \textit{does not require a separate speaker-verification module, an 
enrollment recording, or a manually tuned rejection threshold}.

A further challenge is that, to the best of our knowledge, no large-scale wake-plus-command training 
data is publicly available for the enrollment-free target-speaker ASR task studied in this work.
Learning the relationship between a wake-word speaker and a subsequent command requires examples 
containing both matching and mismatching speaker identities under diverse acoustic conditions. 
However, large-scale recordings with explicit wake-word-to-command speaker relationships are difficult 
to obtain. Moreover, directly concatenating speech segments from different recordings may introduce 
unintended acoustic artifacts that correlate with the training labels, allowing the model to exploit 
superficial differences rather than learn speaker identity. To address this issue, we propose a 
large-scale wake-plus-command data synthesis and simulation pipeline that combines speaker-labelled 
speech with Qwen3-TTS~\cite{hu2026qwen3tts} synthesis and explicit speaker-relation controls. Positive and negative examples are constructed with controls over segment composition, silence intervals, recording conditions, and duration distributions. Together with online acoustic augmentation and a progressive four-stage training strategy, 
this pipeline supports the learning of robust target-speaker recognition and rejection under real-world 
overlapping speech and background noise scenarios.

Our main contributions are summarized as follows:
\begin{itemize}[leftmargin=0pt,itemindent=2.5em]
\item \textbf{WakeCue-ASR: Enrollment-free, wake-word-guided Speech-LLM architecture:} 
We propose a novel enrollment-free Speech-LLM architecture that integrates a text-conditioned speech encoder with an LLM decoder. The speech encoder leverages wake-word text to implicitly capture the speaker characteristics of the wake-word speaker, while the LLM decoder transcribes the target command or rejects non-target speech within a unified autoregressive decoding process. By avoiding speaker 
enrollment and separate verification, it supports a more seamless interaction flow on 
shared devices.

\item \textbf{WakeCue-ASR task tailored training data generation pipeline:} 
We propose and publicly release the first pipeline tailored to this task, which combines 
speaker-labelled speech with TTS synthesis and explicit speaker-relation controls to 
build matched and mismatched wake-word-command pairs while avoiding label-correlated acoustic shortcuts.

\item \textbf{Progressive four-stage Speech-LLM training strategy:}
We develop a staged training procedure for our WakeCue-ASR, spanning text-conditioned 
CTC pretraining, Speech-LLM alignment, joint optimization, and auxiliary speaker supervision, 
to learn robust speaker-selective behavior under overlapping speech and noise.

\item \textbf{Validation on real-world industrial scenarios:}
We validate the proposed WakeCue-ASR on two real-world industrial test sets 
(including the Midea AI Challenge and Baidu internal test sets) and one simulated dataset, 
where it substantially outperforms the strongest enrollment-free baseline, 
offering a practical reference for seamless human-machine interaction.

\end{itemize}

\section{Related Work}
\label{sec:related}

\subsection{Target-Speaker ASR}
SpeakerBeam~\cite{zmolikova2019speakerbeam} and its recognition variants~\cite{delcroix2018single} 
condition a separation or recognition network on an embedding extracted from an enrollment utterance, 
and Kanda et al.~\cite{kanda2019auxiliary} add an auxiliary interference-speaker loss to TS-ASR.
VoiceFilter~\cite{wang2019voicefilter} and personal voice activity detection~\cite{ding2020personal} 
use a d-vector of the enrolled user to suppress other speakers.
With large pretrained recognizers, Whisper~\cite{radford2023whisper} has been adapted to the 
target speaker by prompt tuning with a speaker embedding~\cite{ma2024extending}, and 
SQ-Whisper~\cite{guo2025sqwhisper} queries the Whisper encoder with enrollment features.
LLM-based TS-ASR follows the same principle: Work in~\cite{kim2026target} feed the 
output of a speaker-aware speech encoder into a frozen LLM, and 
Xiaomi-CocktailASR-1~\cite{zhang2026xiaomi} prepends enrollment speech as a 
voice-print prompt and outputs empty text when the target is absent.
All of these methods assume that a clean enrollment of the target speaker is available before the interaction.

\subsection{Anchored and Device-Directed Speech Recognition}
Anchored speech recognition replaces enrollment with the wake-word segment of the current request.
Wang et al.~\cite{wang2019anchored} computed an anchor embedding from the wake word and 
feed it to an attention-based recognizer to suppress background speech, and Raj et al.~\cite{raj2023anchored} 
extended the idea to neural transducers with a cross-attention anchor.
Device-directed speech detection~\cite{mallidi2018device} decides whether an utterance is addressed to 
the device, and SELMA~\cite{wagner2025selma} folds voice-trigger detection, 
device-directedness detection, and ASR into one speech LLM.
Neither line of work selects a target talker in overlapped speech, and the anchored methods 
take the wake segment as audio. Our method takes the wake word as text, which avoids 
extracting an embedding from a short and noisy segment and lets the model learn where 
in the mixture the cue was spoken.

\subsection{Text Cues for Target Speech Processing}
Text has been used as a query for target speaker extraction, 
either as a natural-language description of the target speaker~\cite{hao2024typing} or 
as a keyword from the target's transcript~\cite{li2026detect}.
For recognition, Shi et al.~\cite{shi2024keyword} conditioned an attention-based encoder-decoder 
on a keyword encoder and transcribed the LibriMix talker whose speech contains the keyword.
Our earlier work~\cite{mei2026unified} injected a text cue, an enrollment cue, or both into a 
pretrained speech encoder through cross-attention and trained a CTC recognizer with a rejection label; 
the encoder in this paper follows the same conditioning design.
MT-LLM~\cite{meng2024large} showed that an LLM can transcribe the talker who spoke a given keyword 
in simulated LibriSpeechMix mixtures, as one of several instruction types.
It does not reject inputs in which the cue speaker is absent or the command comes from another speaker, 
and it does not address a wake-word prefix in noisy device recordings.
The present work differs from these methods in three respects: the cue is the wake word that precedes 
the command in the same utterance, rejection is part of the LLM output vocabulary rather than a 
separate classifier or threshold, and the training data are built for the wake-word-then-command 
structure.

\subsection{Multi-Talker ASR and Speech LLMs}

Multi-talker ASR aims to transcribe overlapping speech from multiple speakers. Representative 
approaches include serialized output training (SOT)\cite{kanda2020serialized}, which generates 
transcripts from multiple speakers in a serialized sequence, and its extensions with LLM-based decoders\cite{shi2024advancing}. 
Alternatively, speech separation models such as Conv-TasNet~\cite{luo2019convtasnet} can be 
combined with conventional ASR systems to recognize individual speech streams. Recent end-to-end systems, 
including SoulX-Transcriber~\cite{dai2026soulxtranscriber} and MOSS-Transcribe-Diarize~\cite{moss2026transcribe}, 
further integrate speaker-attributed transcription and diarization. However, these approaches primarily 
focus on recognizing or attributing speech from multiple speakers rather than identifying a designated speaker 
from a wake-word cue. Applying them to wake-word-driven interaction generally requires an additional 
matching step to associate the wake-word occurrence with the corresponding speaker and extract the target 
command. Such post-processing can be sensitive to speaker-attribution inconsistencies and wake-word recognition errors.

In parallel, Speech-LLMs have advanced speech recognition and audio-language understanding by integrating 
speech encoders with pretrained language models. Representative systems include SALMONN~\cite{tang2024salmonn}, 
SLAM-ASR~\cite{ma2024embarrassingly}, and Seed-ASR~\cite{bai2024seedasr}, which explore audio-language alignment, 
efficient ASR adaptation, and contextual information integration. More recently, Qwen3-Omni~\cite{xu2025qwen3omni} 
has extended speech-language modeling to unified multimodal interaction, while Qwen3-ASR~\cite{shi2026qwen3asr} 
provides a dedicated speech recognition framework with multilingual capabilities. Despite these advances, 
general-purpose Speech-LLMs do not explicitly address the association between a wake-word occurrence and 
its speaker for subsequent command recognition and rejection. In contrast, our WakeCue-ASR incorporates 
the wake-word text directly into the acoustic encoding process, enabling target-speaker-conditioned 
recognition and explicit rejection within a unified autoregressive model.

\begin{figure*}[ht]
\centering
\includegraphics[width=0.8\textwidth]{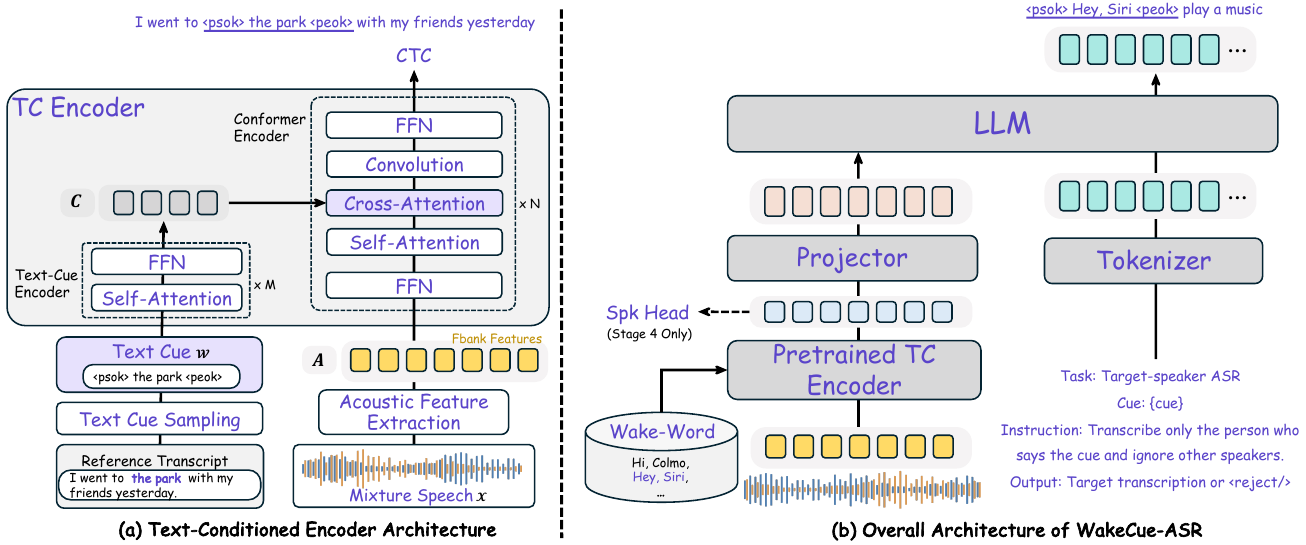}
\caption{Overview of the proposed WakeCue-ASR framework. (a) Architecture and CTC-based pretraining of the text-conditioned (TC) speech encoder, in which contextual representations of the text cue interact with acoustic representations through cross-attention. (b) Overall architecture of WakeCue-ASR, which integrates the pretrained TC encoder, an acoustic projector, and an LLM decoder for enrollment-free target command transcription and non-target speech rejection.}
\label{fig:arch}
\end{figure*}

\section{Proposed WakeCue-ASR Framework}
\label{sec:method}

Let $x$ denote the observed mixture and $w$ the known wake-word text. The objective is to 
recover the command spoken by the speaker who uttered $w$, even when the command overlaps 
with interfering speech. If the wake word is absent from $x$, or if the subsequent command 
is spoken by another speaker, the input must be rejected. The proposed formulation requires 
neither an enrollment utterance nor an externally supplied wake-word boundary.

As shown in Fig.~\ref{fig:arch}(a), the text-cue encoder converts $w$ into contextual lexical representations, which are integrated with acoustic representations through cross-attention in the text-conditioned speech encoder. Fig.~\ref{fig:arch}(b) shows the complete WakeCue-ASR system, comprising TC encoder, an acoustic projector, and an autoregressive LLM decoder. The resulting acoustic sequence is mapped to the input space of a pretrained LLM, which generates either the target command or a rejection decision in one decoding process.

\subsection{Text-Cue Encoder}
Following the text-cue conditioning strategy in~\cite{mei2026unified}, the known wake word 
is first mapped to a contextual token sequence by a lightweight Transformer text encoder:
\begin{equation}
\mathbf{C}=f_{\mathrm{text}}(w)
\label{eq:cue-encoding}
\end{equation}
The resulting sequence preserves token-level lexical information and is used as the conditioning 
sequence for the acoustic encoder. In contrast to an enrollment-based representation, 
it \textit{does not} require a separate recording from the target speaker or an externally determined cue boundary.

\subsection{Text-Conditioned Speech Encoder}
\label{sec:method:encoder}

As illustrated in Fig.~\ref{fig:arch}(a), the text-conditioned (TC) speech encoder is built upon the Conformer architecture~\cite{gulati2020conformer} and incorporates cross-attention modules to enable interactions between acoustic and textual representations. Given the observed waveform $x$ and the text cue $w$, the acoustic input representations and contextual cue representations are obtained as
\begin{equation}
\mathbf{A}=f_{\mathrm{ac}}(x), \qquad
\mathbf{C}=f_{\mathrm{text}}(w)
\end{equation}
where $f_{\mathrm{ac}}(\cdot)$ and $f_{\mathrm{text}}(\cdot)$ denote the acoustic feature extraction and text encoding functions, respectively.

Within the TC encoder, cross-attention is incorporated into the Conformer blocks to condition acoustic representations on the textual cue. Specifically, the intermediate acoustic representation $\mathbf{Q}$ serves as the query, while the contextual cue representation $\mathbf{C}$ serves as both the key and value:
\begin{equation}
\widetilde{\mathbf{Q}}=
\operatorname{LN}\left(
\mathbf{Q}+\operatorname{CA}(\mathbf{Q},\mathbf{C},\mathbf{C})
\right)
\label{eq:cue-cross-attention}
\end{equation}
Here, $\operatorname{LN}(\cdot)$ denotes layer normalization, and $\operatorname{CA}(\cdot,\cdot,\cdot)$ denotes the cross-attention operation.
Through repeated text-acoustic interactions and temporal context modeling, the encoder learns to associate the textual cue with its acoustic realization and propagate the corresponding speaker information throughout the acoustic sequence. The overall encoding process is formulated as
\begin{equation}
\mathbf{H}=f_{\mathrm{TC}}(\mathbf{A},\mathbf{C})
\label{eq:tc-encoder}
\end{equation}
where $\mathbf{H}$ denotes the resulting text-conditioned acoustic representation.

\subsection{Acoustic Projector and LLM Decoder}
\label{sec:method:llm}

To bridge the representation gap between the TC encoder and the pretrained LLM, an acoustic projector is employed to map the text-conditioned acoustic representation $\mathbf{H}$ into the LLM embedding space:
\begin{equation}
\mathbf{Z}=f_{\mathrm{proj}}(\mathbf{H})
\label{eq:projector}
\end{equation}
where $f_{\mathrm{proj}}(\cdot)$ consists of a gated one-dimensional convolutional module for temporal downsampling and a linear projection for dimensional alignment. The projected acoustic embeddings $\mathbf{Z}$ are combined with the task instruction and wake-word text $w$ to construct a multimodal prompt $P(w,\mathbf{Z})$, which serves as the input prefix to the LLM decoder. A simplified English illustration of the task prompt is provided in Fig.~\ref{fig:arch}(b), while the complete prompt is available in our code repository.

To unify target command transcription and non-target speech rejection within a single generative framework, we formulate both tasks as conditional sequence generation. Three special tokens, \psok{}, \peok{}, and \rej{}, are introduced into the LLM vocabulary, where \psok{} and \peok{} delimit the wake-word transcription, and \rej{} indicates rejection. Given the wake-word text $w$ and target command transcript $c$, the target sequence $y$ is constructed according to the input conditions summarized in Table~\ref{tab:target-sequence}.

\begin{table}[t]
\centering
\caption{Target sequence construction under different input conditions, where $w$ and $c$ denote the wake-word text and target command transcript, respectively.}
\label{tab:target-sequence}
\setlength{\tabcolsep}{3pt}
\begin{tabular}{p{0.46\columnwidth}p{0.48\columnwidth}}
\toprule
Condition & Target sequence $y$ \\
\midrule
Target command present & $\psok,w,\peok,c$ \\
Command from another speaker & $\psok,w,\peok,\rej$ \\
Wake-word cue absent & $\rej$ \\
\bottomrule
\end{tabular}
\end{table}

Explicit wake-word transcription provides structured supervision for associating the textual cue with its acoustic realization, while the subsequent token generation determines whether to transcribe the target command or reject the non-target speech.

The LLM decoder is optimized using the teacher-forced autoregressive cross-entropy objective:
\begin{equation}
\mathcal{L}_{\mathrm{CE}}=-\sum_{t=1}^{|y|}\log p_{\theta}(y_t\mid y_{<t},P(w,\mathbf{Z}))
\label{eq:ce}
\end{equation}
where $\theta$ denotes the trainable parameters of the LLM decoder. To enable parameter-efficient adaptation, LoRA~\cite{hu2022lora} is employed while keeping the pretrained LLM weights frozen.

During inference, the LLM generates the output sequence through greedy autoregressive decoding:
\begin{equation}
\hat{y}=f_{\mathrm{LLM}}(P(w,\mathbf{Z}))
\label{eq:llm-decoding}
\end{equation}
The final recognition result $\hat{c}$ is obtained by parsing the generated sequence. If $\hat{y}$ contains \rej{} or no token follows \peok{}, the input is rejected ($\hat{c}=\varnothing$); otherwise, the text following \peok{} is returned as the target command. This unified generation framework enables target command transcription and non-target speech rejection within a single decoding process, without requiring an external speaker verification module or an additional decision threshold.

\subsection{Auxiliary Speaker Supervision}
\label{sec:method:spk}

\begin{figure}[t]
\centering
\includegraphics[width=\columnwidth]{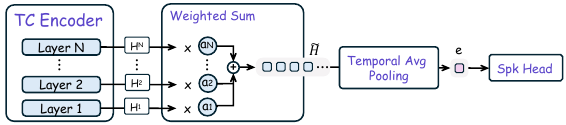}
\caption{Stage~4 auxiliary speaker-classification head in WakeCue-ASR. It aggregates text-conditioned encoder representations across layers using a weighted sum and averages over non-padded frames to predict the wake-word speaker identity. The resulting speaker loss supplements the LLM objective during Stage~4 training.}
\label{fig:spk}
\end{figure}

To enhance the speaker-discriminative capability of the text-conditioned acoustic representations, we introduce an auxiliary speaker classification objective that explicitly supervises the encoder to capture the identity of the wake-word speaker. As illustrated in Fig.~\ref{fig:spk}, the auxiliary speaker head operates on the intermediate representations of the TC encoder, following the weighted layer aggregation and temporal average pooling strategy in~\cite{li2026detect}.

Specifically, the hidden representations $\{\mathbf{H}^{l}\}_{l=1}^{N}$ from the $N$ encoder layers are aggregated through a learnable weighted sum, followed by temporal average pooling over non-padded frames to obtain an utterance-level speaker representation:
\begin{equation}
\mathbf{e}=\operatorname{AvgPool}\left(\sum_{l=1}^{N}a_l\mathbf{H}^{l}\right)
\label{eq:spk-embedding}
\end{equation}
where $N$ denotes the number of encoder layers and $a_l$ is the learnable aggregation weight associated with the $l$-th layer. The resulting speaker representation $\mathbf{e}$ is fed into a linear classification head to predict the identity of the wake-word speaker:
\begin{equation}
\mathbf{o}=f_{\mathrm{spk}}(\mathbf{e})
\label{eq:spk-classification}
\end{equation}
where $\mathbf{o}$ denotes the predicted speaker logits.

Given the ground-truth wake-word speaker identity $s^{w}$, the auxiliary speaker objective is formulated as
\begin{equation}
\mathcal{L}_{\mathrm{spk}}=
\operatorname{CE}(s^{w},\mathbf{o})
+\beta\left(\lVert\mathbf{a}\rVert_{2}-1\right)^{2}
\label{eq:spk-loss}
\end{equation}
where $\mathbf{a}$ denotes the vector of layer aggregation weights and $\beta$ controls the strength of the norm regularization. The auxiliary speaker objective encourages the TC encoder to retain speaker-discriminative information associated with the wake-word speaker rather than relying solely on lexical information. For cue-absent samples, the speaker loss is omitted since no target speaker is defined.

The overall training objective incorporating auxiliary speaker supervision is formulated as
\begin{equation}
\mathcal{L}=\mathcal{L}_{\mathrm{CE}}+\lambda\mathcal{L}_{\mathrm{spk}}
\label{eq:total-loss}
\end{equation}
where $\lambda$ balances the autoregressive sequence generation objective and the auxiliary speaker classification objective. The auxiliary speaker head is employed exclusively during training and discarded at inference, introducing no additional inference overhead.

\section{Training Data Generation and Online Simulation}
\label{sec:data}
\begin{figure*}[t]
\centering
\includegraphics[width=0.7\textwidth]{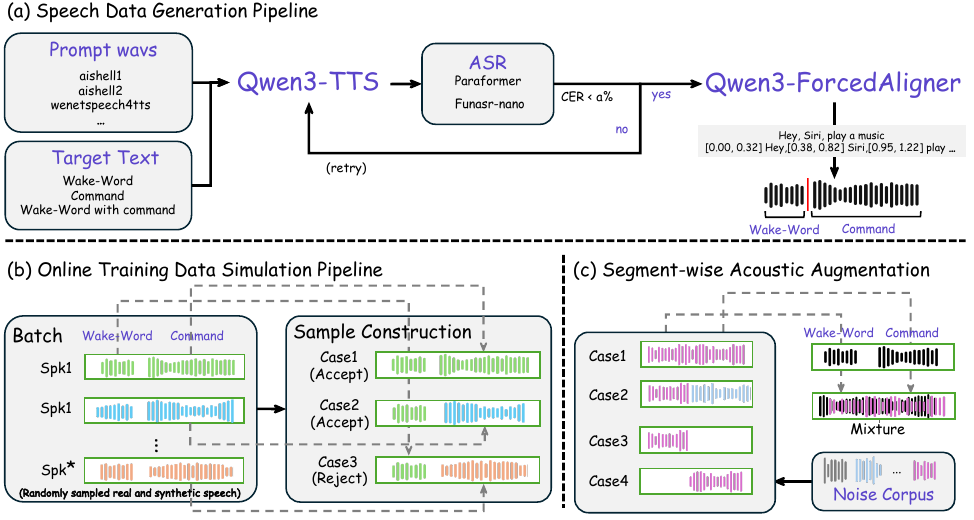}
\caption{Overview of the training-data generation pipeline. (a) Qwen3-TTS synthesizes wake-word-only, command-only, and wake-plus-command utterances from speaker-labelled prompts, followed by ASR-based quality filtering and forced alignment. (b) Online batches contain randomly sampled real and synthesized speech. The same speaker-aware sample-construction procedure is applied to both sources to generate three configurations: wake-word and command segments from the same utterance, different utterances of the same speaker, or different speakers. The first two constitute acceptance samples, while the third constitutes rejection samples. (c) Segment-wise acoustic augmentation introduces background noise and up to two interfering speakers online. Noise can be applied continuously across both segments, independently to each segment, or exclusively to either segment.}
\label{fig:pipeline}
\end{figure*}

The target task requires the model to determine whether the command speaker is the person who 
uttered the wake word. To address the lack of public datasets for this enrollment-free setting 
and prevent the model from exploiting spurious acoustic correlations,
We therefore construct speaker-labelled training examples with both matching and mismatching 
wake-command identities. The data pipeline combines speaker-labelled source speech, 
Qwen3-TTS~\cite{hu2026qwen3tts} synthesis, manifest-based spk-mix composition, and online 
acoustic augmentation, as summarized in Fig.~\ref{fig:pipeline}.

\subsection{Real Speech Sources}
Real training speech is drawn from speaker-labelled AISHELL-1~\cite{bu2017aishell1} and AISHELL-2~\cite{du2018aishell2}, as well as a 2,000-h subset of WenetSpeech~\cite{zhang2022wenetspeech}. WenetSpeech is used in Stages1–3 but excluded from Stage4 because the selected subset lacks the speaker labels required for auxiliary speaker supervision (see Section~\ref{sec:data:stages}).

\subsection{Data Generation Pipeline}

Prompt waveforms for Qwen3-TTS are drawn from AISHELL-1/2/3~\cite{bu2017aishell1,du2018aishell2,shi2021aishell3} and the Premium subset of WenetSpeech4TTS~\cite{ma2024wenetspeech4tts}. AISHELL-3 and WenetSpeech4TTS are used exclusively as TTS prompt sources, and their original recordings are not directly included in the training corpus. Original speaker IDs are retained for AISHELL-1/2/3, whereas pseudo-speaker labels are assigned to WenetSpeech4TTS prompt waveforms. Each synthesized utterance inherits the speaker label associated with its prompt, which is stored in the data manifest for speaker-aware sample construction and auxiliary speaker supervision.

As illustrated in Fig.\ref{fig:pipeline}(a), Qwen3-TTS synthesizes three types of utterances: wake-word-only, command-only, and wake-plus-command utterances. To ensure the transcription fidelity of synthesized speech, an ASR-based quality filter is applied by comparing the recognized text with the reference transcript. As substitution errors dominate the ASR hypotheses for synthesized speech, we adopt a CER threshold of 20\% to balance data quality and sample retention. Utterances exceeding this threshold are resynthesized using an alternative prompt waveform from the same speaker. Wake-plus-command utterances are further processed using Qwen3-ForcedAligner\cite{shi2026qwen3asr} to obtain temporal boundaries between the wake word and the command, enabling subsequent segment-level sample construction. Separately synthesized wake-word and command utterances from the same speaker are also paired offline to increase the diversity of training samples.

As shown in Fig.~\ref{fig:pipeline}(b), we employ an online speaker-aware sample construction strategy to dynamically generate acceptance and rejection examples from real and synthesized utterances randomly sampled within the current training batch. Three configurations are considered: (1) the wake word and command originate from the same utterance, preserving natural acoustic continuity; (2) the two segments originate from different utterances of the same speaker, introducing variations in acoustic conditions while preserving speaker consistency; and (3) the two segments originate from different speakers, simulating scenarios in which the wake-word speaker does not issue the subsequent command. The first two configurations are treated as acceptance examples, whereas the third is treated as a rejection example. In addition, utterances that do not contain the specified wake word are included as rejection examples, enabling the model to reject inputs in the absence of the wake-word cue. Importantly, introducing cross-utterance concatenation into acceptance examples is designed to discourage the model from exploiting acoustic discontinuities at segment boundaries as shortcuts for rejection, encouraging it to rely instead on speaker consistency between the wake word and the command.

Following sample construction, segment-wise acoustic augmentation is performed online, as illustrated in Fig.~\ref{fig:pipeline}(c). Four noise configurations are considered: continuous noise across both segments, different noise across the two segments, noise applied only to the wake-word segment, and noise applied only to the command segment. These configurations simulate both consistent and changing acoustic environments during voice interaction. Interfering speech is introduced using a similar segment-wise strategy, with up to two utterances from distinct non-target speakers randomly sampled from the current training batch. The dynamic combination of speaker-aware sample construction and acoustic augmentation increases training diversity across iterations while preserving the acceptance or rejection labels determined by the speaker consistency between the original wake-word and command segments and the presence of the specified wake-word cue.
\subsection{Four-Stage Speech-LLM Training}
\label{sec:data:stages}

\begin{table*}[t]
\centering
\caption{Trainable modules and parameter counts across the four training stages of WakeCue-ASR. TC enc., CTC, Proj., and Spk. head denote the text-conditioned speech encoder, CTC head, acoustic projector, and auxiliary speaker-classification head, respectively, as illustrated in Fig.~\ref{fig:arch}. \checkmark~indicates trainable parameters, frozen'' indicates frozen parameters, and --'' indicates the absence of the corresponding module. Only the LoRA adapters are trainable within the LLM.}
\label{tab:stages}
\setlength{\tabcolsep}{6pt}
\begin{tabular}{ccccccc}
\toprule
Stage & TC enc. & CTC & Proj.+LLM (LoRA) & Spk. head & Total Param. (M) & Trainable Param. (M) \\
\midrule
1 & \checkmark & \checkmark & -- & -- & 37.18 & 37.18 \\
2 & frozen & -- & \checkmark & -- & 577.41  & 48.60 \\
3 & \checkmark & -- & \checkmark & -- & 577.41  & 83.62 \\
4 & \checkmark & -- & \checkmark & \checkmark & 579.17 & 85.38 \\
\bottomrule
\end{tabular}
\end{table*}

\begin{table}[t]
\centering
\caption{Training data composition and nominal corpus duration at each stage.
The reported durations exclude additional training exposure introduced by
online sample construction and acoustic augmentation.}
\label{tab:stage_data}
\footnotesize
\setlength{\tabcolsep}{3pt}
\begin{tabular}{@{}c p{0.62\columnwidth} c@{}}
\toprule
Stage & Datasets & Duration (hrs)\\
\midrule
1 & AISHELL-1/2 + Wenet & 3K \\
2 & AISHELL-1/2 + Wenet + TTS + TTS-Concat & 101K \\
3 & AISHELL-1/2 + Wenet + TTS + TTS-Concat & 101K \\
4 & AISHELL-1/2 + TTS + TTS-Concat & 30K \\
\bottomrule
\end{tabular}

\par\vspace{4pt}
\noindent
\parbox{\columnwidth}{%
\footnotesize
\textit{Notes:} Wenet denotes the 2K-hrs subset of WenetSpeech.
TTS denotes speech directly synthesized with Qwen3-TTS, while
TTS-Concat denotes samples constructed offline by concatenating
separately synthesized wake-word and command utterances from
the same speaker.
Stages~2 and~3 share the same training corpus.
Stage~4 excludes WenetSpeech due to the absence of speaker labels
and uses a speaker-labelled subset comprising 6,855 speaker classes:
2,330 real speakers from AISHELL-1/2 and 4,525 TTS voices generated
from AISHELL-1/2/3 and WenetSpeech4TTS prompts, with at least
20 utterances per speaker.
}
\end{table}

We adopt a four-stage training strategy to learn text-conditioned speech representations, align them with the pretrained LLM, and progressively adapt the model to wake-word-guided target-speaker ASR. The trainable modules and training data at each stage are summarized in Tables~\ref{tab:stages} and~\ref{tab:stage_data}, respectively.

\textbf{Stage 1: Text-conditioned encoder pretraining.}
As shown in Fig.\ref{fig:arch}(a), the TC encoder is pretrained with the CTC objective\cite{graves2006ctc}. For each positive sample, a contiguous span of 2–5 characters is randomly selected from the reference transcript as the text cue. The cue boundaries in the target sequence are marked with $\psok$ and $\peok$. Following the negative-text sampling strategy in~\cite{mei2026unified}, negative samples use text cues that do not occur in the reference transcript and are supervised with the rejection token. Rejection supervision at this stage is therefore based solely on cue absence, without introducing speaker switching between speech segments. This stage establishes text-conditioned speech modeling and cue-validity discrimination prior to integration with the LLM.

\textbf{Stage 2: Speech–LLM alignment.}
As shown in Fig.~\ref{fig:arch}(b), we remove the CTC head and connect the pretrained TC encoder to the LLM through an acoustic projector. The TC encoder remains frozen, while only the acoustic projector and LLM LoRA parameters are optimized for autoregressive transcription and rejection. From this stage onward, training adopts the wake-word–command formulation and the online data simulation strategy described above, incorporating both wake-word-absent and speaker-mismatched rejection samples. This stage aligns the text-conditioned speech representations with the LLM embedding space while adapting the decoder to the target task.

\textbf{Stage 3: End-to-end joint optimization.}
We unfreeze the TC encoder and jointly optimize it with the acoustic projector and LLM LoRA parameters under the same autoregressive supervision. This stage enables joint adaptation of the speech representations and the LLM decoder to wake-word-guided target-speaker ASR.

\textbf{Stage 4: Auxiliary speaker supervision.}
We introduce an auxiliary speaker-classification head to strengthen speaker-discriminative information in the TC encoder representations. Training is conducted on the speaker-labelled subset comprising 6,855 speaker classes. All previously trainable modules and the auxiliary head are jointly optimized under both task-specific and speaker-classification supervision. This additional supervision encourages the TC encoder to preserve speaker identity information relevant to speaker-consistency modeling between the wake word and the subsequent command.
\section{Experimental Setup}
\label{sec:setup}

\subsection{Evaluation Data}
We evaluate the proposed system on \textit{midea-test}, a smart-home wake-plus-command dataset provided 
by the Midea AI challenge~\footnote{\url{https://ai.midea.com/\#/match-detail}}. 
It contains 1,838 utterances recorded under challenging household acoustic conditions, including overlapping 
speech and interference from air conditioners, household appliances, music, singing, background speech, and 
other environmental sounds. The SNR ranges from $-5$ to $5$\,dB, the speaker-overlap ratio ranges from 0 to 100\%, 
and at most two speakers are active simultaneously.

For additional real-world validation, we use an internal dataset, \textit{baidu-test}, collected from Baidu’s commercial voice assistant scenario. It contains 2,991 real recordings collected under varying air-conditioner fan levels, background interference conditions, recording distances, and source directions. The background interference may include television audio, singing, and speech from other speakers, in addition to conventional environmental noise. The dataset is divided into ten subsets to evaluate robustness under diverse household acoustic conditions and far-field recording configurations.

We further evaluate the generality of text-cue conditioning on the TS-AISHELL target-speaker test sets~\cite{mei2026unified}\footnote{\url{https://huggingface.co/datasets/YuCeong-May/Unified-TS-ASR-Eval}}. 
The cue is a span of 2--5 characters selected from the beginning of the target speaker's sentence to 
simulate a wake word preceding the command, and each mixture contains an interfering talker. 
The benchmark comprises five subsets: AISHELL-1, the iOS, Android, and microphone channels of AISHELL-2, 
and a mixed-channel set. Each subset contains 6000 unique mixtures, which are evaluated with 
four cue lengths to produce 120\,000 text-cue evaluation instances in total.

\subsection{Evaluation Metrics}
\label{sec:setup:metrics}

We adopt different evaluation metrics according to the characteristics of each test set. For \textit{midea-test}, we report the command CER on positive samples (pos-CER), the false-rejection rate (FR), and the rejection rate (RR). A rejected positive sample is treated as deleting all reference characters when computing pos-CER. FR is defined as the proportion of positive samples with empty outputs, whereas RR is defined as the proportion of negative samples with empty outputs. To jointly assess recognition and rejection performance on \textit{midea-test}, we further adopt a unified score:
\begin{equation}
U=\frac{100-\mathrm{pos\text{-}CER}+\mathrm{RR}}{2}
\end{equation}
where pos-CER and RR are expressed as percentages. This score assigns equal weight to positive-sample recognition and negative-sample rejection.

For \textit{baidu-test}, we report CER and FR; for TS-AISHELL, we report CER only.
All systems are evaluated with a batch size of 1, and LLM-based systems employ autoregressive greedy decoding.

\begin{table*}[t]
\centering
\caption{Results on \textnormal{\textit{midea-test}}. All metric values are percentages, and $U$ is computed from unrounded rates before being rounded to two decimal places. Qwen3-ASR components use Qwen3-ASR-1.7B. The B2 cosine-similarity threshold $\tau$ is selected by scanning this same test set.}
\label{tab:main}
\setlength{\tabcolsep}{4pt}
\begin{tabular}{>{\centering\arraybackslash}p{0.06\textwidth}p{0.27\textwidth}cccc>{\raggedright\arraybackslash}p{0.25\textwidth}}
\toprule
ID & System & pos-CER$\downarrow$ & FR$\downarrow$ & RR$\uparrow$ & $U\uparrow$ & Notes \\
\midrule
B1 & Qwen3-ASR-1.7B~\cite{shi2026qwen3asr} & 32.21 & 0.00 & 0.00 & 33.89 & Direct ASR \\
B2 & CAM++~\cite{wang2023campp} + Qwen3-ASR~\cite{shi2026qwen3asr} & 46.26 & 27.05 & 89.66 & 71.70 & Speaker Gating \& ASR; $\tau=0.28$ \\
B3 & Qwen2.5-Omni-7B~\cite{xu2025qwen25omni} & 44.04 & 28.01 & 21.73 & 38.85 & Audio-LLM Prompting \\
B4 & MOSS-Transcribe-Diarize~\cite{moss2026transcribe} & 51.44 & 35.92 & 93.46 & 71.01 & Multi-talker ASR \& match \\
B5 & WeSep~\cite{wang2024wesep} + Qwen3-ASR~\cite{shi2026qwen3asr} & 73.99 & 1.25 & 29.11 & 27.56 & Speech separation \& ASR \\
B6 & CocktailASR-1~\cite{zhang2026xiaomi} & 29.19 & 9.60 & 66.67 & 68.74 & Wake as ref. \\
\midrule
S1 & WakeCue-ASR, Stage 1 (CTC) & 43.47 & 45.82 & \textbf{98.10} & 77.31 & Text-Cue Encoder \\
S2 & WakeCue-ASR, Stage 2 & 29.15 & 20.67 & 93.25 & 82.05 & LLM Alignment \\
S3 & WakeCue-ASR, Stage 3 & 20.47 & 13.86 & 90.72 & 85.12 & Joint Training \\
S4 & WakeCue-ASR, Stage 4 & \textbf{16.15} & \textbf{9.46} & 91.56 & \textbf{87.70} & Speaker Supervision \\
\bottomrule
\end{tabular}
\end{table*}

\subsection{Baselines}
\label{sec:setup:baselines}
We compare with six baselines that require no pre-enrolled speaker recordings. They are built from public pretrained models and cover the conventional designs discussed in Section~\ref{sec:related}.
None of them is trained on the target task.

\emph{B1: direct speech recognition.}
Qwen3-ASR~\cite{shi2026qwen3asr} directly transcribes the command segment 
without using the wake segment as a speaker reference or applying any explicit speaker verification or rejection mechanism. This baseline serves as a reference for evaluating recognition performance without target-speaker constraints.

\emph{B2: speaker-gated cascade.}
CAM++\cite{wang2023campp} embeds the wake segment and the command segment, and Qwen3-ASR\cite{shi2026qwen3asr} transcribes the command only if the cosine similarity of the two embeddings exceeds a threshold; otherwise, the input is rejected.
The threshold is selected by sweeping over candidate values on \textit{midea-test} to maximize the unified score.

\emph{B3: zero-shot audio-LLM prompting.}
We prompt Qwen2.5-Omni-7B~\cite{xu2025qwen25omni} with both textual and speaker cues.
The wake segment is provided first as a reference recording of the target speaker, followed by the command segment as the audio to be transcribed.
The Chinese instruction specifies the wake word and instructs the model to transcribe only the speech of the reference speaker in the command segment while ignoring other speakers.
Decoding is greedy with the talker module disabled.

\emph{B4: multi-talker transcription with text matching.}
MOSS-Transcribe-Diarize~\cite{moss2026transcribe} transcribes the wake-plus-command audio into speaker-attributed segments, with the wake word given as a hotword in its documented prompt.
The speaker whose time-ordered transcript best matches the wake word is taken as the target, and the output is that speaker's text after the wake word.

\emph{B5: target speaker extraction followed by recognition.}
The released WeSep~\cite{wang2024wesep} BSRNN extractor with an ECAPA-TDNN speaker encoder, trained on VoxCeleb1 mixtures, extracts the target speaker from the concatenated wake-plus-command audio, with the wake segment as the enrollment.
Qwen3-ASR transcribes the extracted signal with the wake word as context.

\emph{B6: enrollment-prompted LLM.} \mbox{Xiaomi-CocktailASR-1}~\cite{zhang2026xiaomi} is a target-speaker speech LLM with a data2vec\,2.0 audio encoder and a Qwen3 decoder of about 8\,B parameters.
It prepends a 1--4\,s reference clip and 1\,s of silence to the target audio and outputs the reference speaker's transcript, or empty text when that speaker is absent.
Following the anchored approach, we use the wake segment as the reference clip and the command segment as the target audio, and decode with the released prompt and settings; an empty output counts as a rejection.

B1, B2, and B5 use Qwen3-ASR-1.7B.
B4 and B5 transcribe audio containing the wake word, so the command is extracted as the text following the matched wake word in the target transcript. The input is rejected if no text follows the matched wake word.
A fixed matching rule is applied across both baselines: the transcript and wake word are romanized using toneless pinyin for Chinese characters and lowercase letters otherwise. A match is accepted if the edit distance between the wake word and the best-matching substring does not exceed half the length of the romanized wake word.
If no wake-word match is found, B4 rejects the input, whereas B5 retains the entire transcript as the command.

\subsection{Implementation Details}
All four training stages use the Adam optimizer, gradient clipping with a maximum norm of 5, gradient accumulation over 16 steps, bfloat16 mixed-precision training, and DeepSpeed ZeRO stage1 on four NVIDIA A100 40,GB GPUs.
The TC encoder comprises a nine-block Conformer speech encoder and a four-layer Transformer-based text-cue encoder, both trained from scratch in Stage1.
We employ Qwen2.5-0.5B as the LLM decoder, keeping its pretrained parameters frozen and updating only the LoRA adapters on the LLM side.
The acoustic projector consists of a gated one-dimensional convolution followed by a linear projection, with a temporal downsampling factor of two.
LoRA is configured with a rank of 64, $\alpha=128$, and a dropout rate of 0.05.

During model training, Stage~1 uses a peak learning rate of $1\times10^{-3}$ and is trained for 20 epochs. 
Stage~2 uses $5\times10^{-4}$ with 500 warm-up steps and is trained for 10 epochs. 
Stage~3 uses $5\times10^{-4}$ with 3000 warm-up steps, and the Stage~4 continuation uses
$2\times10^{-4}$ with 500 warm-up steps; both are trained for 10 epochs. Dynamic 
batches contain at most 25\,000 frames in Stage~1, 40\,000 frames in Stage~2, and 34\,000 frames 
with at most 28 utterances in Stages~3--4. Stage~1 is decoded with CTC greedy search, 
whereas Stages~2--4 use autoregressive greedy decoding with the LLM.

\begin{figure}[t]
\centering
\includegraphics[width=\columnwidth]{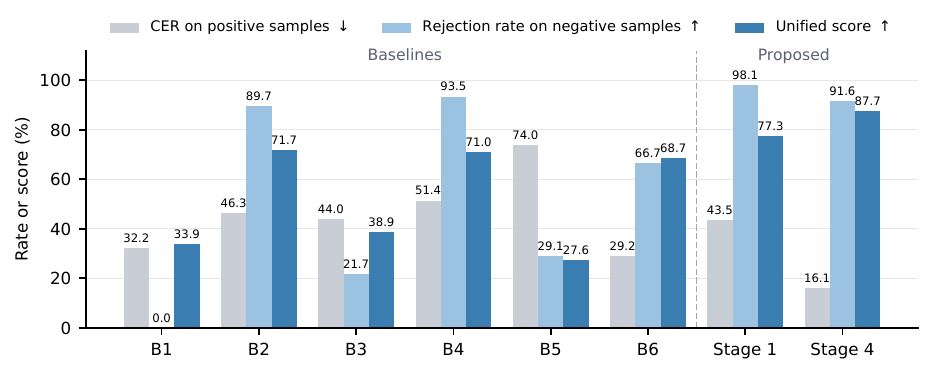}
\caption{pos-CER, RR, and unified score $U$ of the six baselines and of the proposed model after Stage~1 and Stage~4 on \textnormal{\textit{midea-test}}. B2 is shown at its best threshold.}
\label{fig:compare}
\end{figure}

\begin{figure}[t]
\centering
\includegraphics[width=\columnwidth]{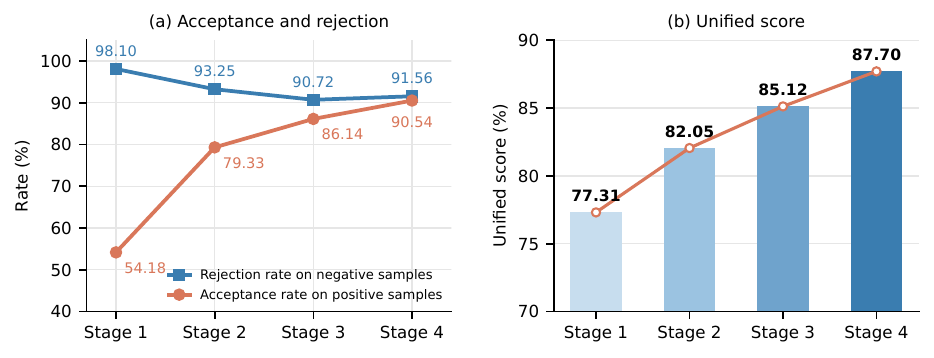}
\caption{Effect of the training stages on \textnormal{\textit{midea-test}}. (a) Fraction of positives accepted and rejection rate on negatives. (b) Unified score $U$.}
\label{fig:stages}
\end{figure}

\begin{figure}[t]
\centering
\includegraphics[width=\columnwidth]{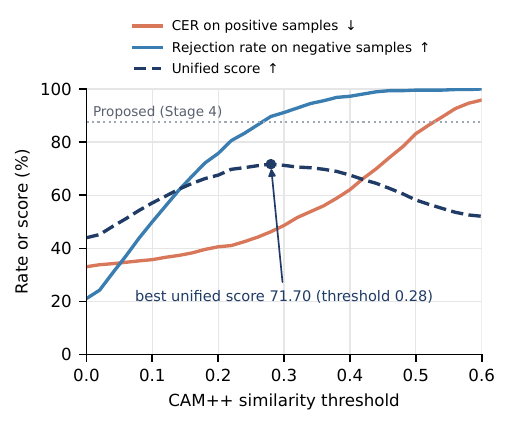}
\caption{Threshold scan of the speaker-gated cascade B2 on \textnormal{\textit{midea-test}}. 
The circular marker on the dotted line marks the unified score of the proposed WakeCue-ASR model 
after Stage~4.}
\label{fig:b1scan}
\end{figure}
\section{Results and Discussions}
\label{sec:results}

\subsection{Comparison with Baselines}
Table~\ref{tab:main} compares all systems on \textit{midea-test}, and Fig.~\ref{fig:compare} 
shows pos-CER, RR, and $U$ of the baselines next to the first and the last training stage of WakeCue-ASR.

The direct speech recognition system B1 achieves 32.21\% pos-CER, but its RR is 0\%. This result shows that a strong general-purpose recognizer can transcribe command speech without target-speaker conditioning, but cannot determine whether the command was issued by the wake-word speaker.
Among the baselines, B2 obtains the highest unified score, 71.70\%, at a cosine-similarity threshold of $\tau=0.28$, where $\tau$ is the minimum similarity between the CAM++ embeddings of the wake and command segments required to pass the speaker gate. At this operating point, it has 46.26\% pos-CER, 27.05\% FR, and 89.66\% RR. The high RR confirms that speaker gating removes many non-target inputs, whereas the simultaneous increase in FR shows that the short, noisy wake segment is an unreliable verification reference for a substantial fraction of genuine users. Thus, the cascade improves rejection by moving to a conservative operating point, but it does not resolve the conflict between target selection and command recognition. Because $\tau$ is selected by scanning \textit{midea-test}, this score is the best operating point on the evaluation set rather than a threshold-independent result. Fig.~\ref{fig:b1scan} shows the corresponding trade-off between pos-CER and RR.

The remaining baselines expose different failure modes of cascaded or prompt-based target selection. B4 has the highest RR (93.46\%) but also a 35.92\% FR and 51.44\% pos-CER. In 330 positive utterances (24.2\%), it assigns the wake word and the following command to different speaker labels, illustrating how diarization errors can disrupt cue matching. Its high RR therefore partly comes from rejecting samples with inconsistent speaker attribution, which explains why strong negative rejection does not translate into reliable positive recognition. B5 rejects few positive utterances (FR 1.25\%) but also rejects only 29.11\% of negatives; its 73.99\% pos-CER is consistent with the high 73.68\% CER on accepted positives. The extractor consequently preserves speech often enough to avoid rejection, but the residual interfering speech remains in the signal and corrupts transcription. B3 obtains a unified score of 38.85\%, with 44.04\% pos-CER, 28.01\% FR, and 21.73\% RR; its CER on accepted positive samples is 24.09\%. Its zero-shot prompt does not provide a reliable mechanism for associating the wake word with one talker or for rejecting the other talker. B6 achieves the lowest pos-CER among the baselines (29.19\%) and an FR of 9.60\%, but its RR is 66.67\%. This indicates that an enrollment-prompted decoder can improve recognition of accepted target speech, while still lacking a sufficiently reliable rejection mechanism when the wake segment is short and noisy.

The proposed WakeCue-ASR system improves across the curriculum. 
It reaches a unified score of 77.31\% at Stage~1 and 87.70\% at Stage~4, where pos-CER is 16.15\%, FR is 9.46\%, and RR is 91.56\%. Stage~1 already exceeds the best baseline in $U$, indicating that text-conditioned acoustic encoding can learn target-speaker selectivity before LLM alignment. Stage~4 further lowers both transcription and false-rejection errors while retaining a high RR, showing that the auxiliary speaker objective improves the balance between accepting genuine commands and rejecting non-target speech. Relative to B2 at its test-set-selected threshold, Stage~4 reduces pos-CER from 46.26\% to 16.15\% and FR from 27.05\% to 9.46\%, increases RR from 89.66\% to 91.56\%, and improves $U$ from 71.70\% to 87.70\%. Unlike the cascaded baselines, the proposed model produces the transcription or rejection directly, so this operating behavior is learned from the task objective rather than selected by a separate verification threshold or transcript-matching rule.

\begin{table}[t]
\centering
\caption{Stage-wise breakdown on \textit{midea-test}.
Acc. denotes the percentage and number of accepted positive utterances.
CER$_{\text{acc}}$ and SER$_{\text{acc}}$ denote the character error rate
and sentence error rate computed on accepted positives only, respectively.}
\label{tab:stagebreak}
\setlength{\tabcolsep}{3.5pt}
\begin{tabular}{cccccc}
\toprule
Stage & Acc. (\% / utt) & CER$_{\text{acc}}$ & SER$_{\text{acc}}$ & pos-CER & RR \\
\midrule
1 & 54.18 / 739  & 21.81 & 57.92 & 43.47 & 98.10 \\
2 & 79.33 / 1082 & 12.49 & 25.51 & 29.15 & 93.25 \\
3 & 86.14 / 1175 & 7.71  & 13.96 & 20.47 & 90.72 \\
4 & 90.54 / 1235 & 6.85  & 10.53 & 16.15 & 91.56 \\
\bottomrule
\end{tabular}
\end{table}

\subsection{Contribution of Each WakeCue-ASR Training Stage}
Table~\ref{tab:stagebreak} and Fig.~\ref{fig:stages} summarize positive-sample acceptance, recognition quality on accepted positives, and rejection of negative samples across the four training stages. Stage~1 is conservative: it accepts 54.18\% of positive samples and achieves a 98.10\% RR, while $CER_{\text{acc}}$ and $SER_{\text{acc}}$, computed on accepted positive samples, are 21.81\% and 57.92\%, respectively. After the frozen TC encoder is aligned with the LLM in Stage~2, acceptance rises to 79.33\% and $CER_{\text{acc}}$ and $SER_{\text{acc}}$ fall to 12.49\% and 25.51\%; this indicates improved command completion, although RR decreases to 93.25\%. Joint optimization in Stage~3 further raises acceptance to 86.14\% and lowers $CER_{\text{acc}}$ and $SER_{\text{acc}}$ to 7.71\% and 13.96\%, while RR declines to 90.72\%, reflecting a temporary shift toward accepting more ambiguous inputs. With auxiliary speaker supervision, Stage~4 improves both acceptance and rejection: acceptance reaches 90.54\%, RR rises to 91.56\%, and $CER_{\text{acc}}$ and $SER_{\text{acc}}$ decrease to 6.85\% and 10.53\%. Overall, the curriculum reduces pos-CER from 43.47\% to 16.15\% and increases the unified score from 77.31\% to 87.70\%, suggesting that speaker supervision helps mitigate the recognition--rejection trade-off.

\begin{table}[t]
\centering
\caption{Evaluation results on \textit{baidu-test}.Subsets N1--N5 and I1--I5 represent scenarios without and with external speech interference, respectively. AC fan level 0 denotes standby mode. Dist. denotes the speaker-to-device distance, Interf. indicates
the presence of external speech interference, and $N$ denotes the number of recordings. All error rates are percentages.}
\label{tab:baidu}
\setlength{\tabcolsep}{4pt}
\begin{tabular}{ccccrrr}
\toprule
Subset & AC fan level & Dist. (m) & Interf. & $N$ & CER$\downarrow$ & FR$\downarrow$ \\
\midrule
N1 & 0 & 5 & \multirow{5}{*}{No} & 298 & 0.18 & 0.00 \\
N2 & 3 & 3 & & 300 & 2.72 & 0.00 \\
N3 & 5 & 5 & & 299 & 0.86 & 0.00 \\
N4 & 6 & 5 & & 298 & 2.89 & 0.00 \\
N5 & 7 & 4 & & 300 & 5.14 & 0.00 \\
\midrule
I1 & 0 & 3 & \multirow{5}{*}{Yes} & 300 & 7.41  & 0.00 \\
I2 & 3 & 3 & & 300 & 10.58 & 1.00 \\
I3 & 5 & 3 & & 300 & 8.01  & 0.00 \\
I4 & 5 & 5 & & 299 & 6.14  & 1.00 \\
I5 & 6 & 5 & & 297 & 8.97  & 1.01 \\
\midrule
\multicolumn{4}{c}{Overall} & 2991 & 5.44 & 0.30 \\
\bottomrule
\end{tabular}
\end{table}

\subsection{Validation on \textit{baidu-test}}
We further evaluate WakeCue-ASR on \textit{baidu-test} without task-specific adaptation.
As shown in Table~\ref{tab:baidu}, for N1–N5 without external speech interference, the CER remains below 3\% except at fan level 7 (5.14\%), with no false rejections.
Under external speech interference (I1–I5), the CER increases to 6.14–10.58\%, whereas the FR remains below 1.02\%, suggesting that interfering speech primarily affects transcription accuracy rather than target acceptance.
Across all 2,991 recordings, WakeCue-ASR achieves a CER of 5.44\% and an FR of 0.30\%, demonstrating robustness in target command recognition across diverse recording conditions. Since \textit{baidu-test} contains only positive samples, RR is not reported.

\begin{figure}[t]
\centering
\includegraphics[width=\columnwidth]{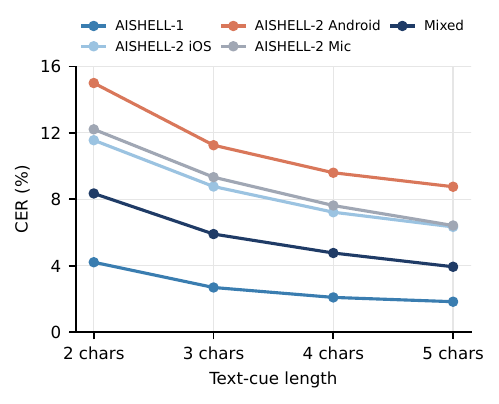}
\caption{Effect of text-cue length on CER across five TS-AISHELL subsets, using sentence-initial text cues (6,000 evaluation instances per point).}
\label{fig:cuelen}
\end{figure}

\subsection{Arbitrary Text Cues Generalization}
The text-conditioning mechanism allows WakeCue-ASR to operate beyond predefined wake words.
We evaluate its generalization on TS-AISHELL using the first 2–5 characters of the target speaker’s sentence as text cues.
As shown in Fig.~\ref{fig:cuelen}, CER consistently decreases with increasing cue length across all five subsets, from 4.21\% to 1.83\% on AISHELL-1 and from 15.00\% to 8.76\% on AISHELL-2 Android.
Longer cues provide greater lexical context for target-speaker localization, with particularly pronounced improvements on AISHELL-2.
Across all 120,000 evaluation instances, the overall CER is 6.96\%, increasing to 9.83\% when the cue span is excluded from both the reference and hypothesis.
Notably, even with two-character cues, the CER remains at or below 15\% across all subsets.
These results demonstrate that WakeCue-ASR can generalize beyond predefined wake words to arbitrary sentence-initial text cues, extending its applicability to text-guided target-speaker ASR.

\section{Conclusion}
\label{sec:conclusion}
In this work, we presented WakeCue-ASR, an enrollment-free Speech-LLM for wake-word-guided target-speaker ASR. By conditioning acoustic representations on wake-word text through cross-attention, the proposed framework integrates target-speaker selection, command transcription, and non-target rejection into a single decoding process. The four-stage training strategy and speaker-aware data simulation enable the model to learn these capabilities without speaker enrollment, explicit wake-word segmentation, or a separate speaker verification module.

Experimental results demonstrate that WakeCue-ASR outperforms existing enrollment-free baselines in balancing target command recognition and non-target rejection. Auxiliary speaker supervision further improves target-speaker discrimination and mitigates the recognition–rejection trade-off. Evaluations on independent real-world recordings demonstrate robustness under diverse acoustic conditions, while experiments on TS-AISHELL verify the generalization capability of the proposed framework beyond predefined wake words to arbitrary sentence-initial text cues.

By unifying target-speaker selection, recognition, and rejection, WakeCue-ASR provides a practical solution for seamless human–machine interaction in complex acoustic environments. Given the availability of real-world industrial evaluation data, this study focuses on Mandarin Chinese. Future work will extend the framework to English and other languages and investigate its multilingual generalization across diverse real-world interaction scenarios.

\section*{Acknowledgment}
This work was supported by the Natural Science Foundation of Shanghai (Grant No. 25ZR1401277) and the National Natural Science Foundation of China (Grant No. 62071302).

\bibliographystyle{IEEEtran}
\bibliography{refs}

\end{document}